\documentclass[10pt,conference]{IEEEtran}
\IEEEoverridecommandlockouts
\usepackage{eso-pic}
\usepackage{pgfplots}
\pgfplotsset{compat=1.17}
\usepackage{pifont}
\usepackage{booktabs}
\newcommand{\cmark}{\ding{51}}
\newcommand{\xmark}{\ding{55}}
\usepackage{amsmath,amssymb,amsfonts}
\usepackage{algorithmic}
\usepackage{graphicx}
\usepackage{textcomp}
\usepackage{xcolor}
\usepackage{cite}
\usepackage{xurl}
\usepackage[hidelinks]{hyperref}
\def\BibTeX{{\rm B\kern-.05em{\sc i\kern-.025em b}\kern-.08em
    T\kern-.1667em\lower.7ex\hbox{E}\kern-.125emX}}

\makeatletter
\newcommand{\linebreakand}{%
  \end{@IEEEauthorhalign}
  \hfill\mbox{}\par
  \mbox{}\hfill\begin{@IEEEauthorhalign}
}
\makeatother

\begin{document}

\AddToShipoutPictureBG*{%
  \AtPageLowerLeft{%
    \raisebox{0.3in}{\makebox[\paperwidth][c]{%
      \parbox[b]{\textwidth}{\centering\footnotesize
        Accepted at the 49th IEEE/ACM International Conference on
        Software Engineering (ICSE 2027).\\
        \copyright~2027 IEEE. Personal use of this material is permitted.
        Permission from IEEE must be obtained for all other uses, in any
        current or future media, including reprinting/republishing this
        material for advertising or promotional purposes, creating new
        collective works, for resale or redistribution to servers or lists,
        or reuse of any copyrighted component of this work in other works.%
      }}}}}

\title{Agent Smith in Action: Internet-scale, Automated Security Analysis of MCP Tools}

\author{%
  \IEEEauthorblockN{%
    Mingming Zha\IEEEauthorrefmark{1},
    Ying Li\IEEEauthorrefmark{2},
    Yuan Tian\IEEEauthorrefmark{2},
    XiaoFeng Wang\IEEEauthorrefmark{3}%
  }
  \IEEEauthorblockA{%
    \IEEEauthorrefmark{1}\,Indiana University Bloomington, USA\\
    \IEEEauthorrefmark{2}\,University of California, Los Angeles, USA\\
    \IEEEauthorrefmark{3}\,Nanyang Technological University, Singapore\\
    mzha@iu.edu,\quad yinglee@ucla.edu,\quad
    yuant@ucla.edu,\quad xiaofeng.wang@ntu.edu.sg
  }
}

\maketitle

\begin{abstract}
The Model Context Protocol (MCP) has rapidly emerged as a common interface through which AI applications and agents connect to external tools, raising a software-engineering challenge for security analysis: independently developed tools are often composed at run time by a large language model (LLM) or a user through files, shared state, parameters, and conversational context, allowing exploitable tool chains to arise even when each individual tool appears benign. These implicit cross-tool dependencies are absent from conventional source-level call graphs, limiting the effectiveness of existing rule-based and interprocedural analyzers for ecosystem-scale MCP auditing. We present SCAN-MCP, a tool-chain-aware framework for detecting security flaws in MCP servers. SCAN-MCP combines LLM-assisted tool-handler mapping, Tree-sitter-based static slicing, semantic source-and-sink inference, inter-tool dependency graph construction, and deterministic structural and type checks. 
The LLM proposes mappings, flows, and proof-of-concept sketches. Source-level and structural checks screen proposals, and analysts manually validate reports from the sampled servers in an isolated sandbox. From a corpus of 10,121 MCP servers with 90,200 discovered tools, SCAN-MCP produced 27,123 reported findings, including 8,422 tool-chain reports (31.05\%). On a language- and marketplace-stratified subset of 100 of these servers, we confirmed 105 of 130 single-tool reports and 19 of 20 tool-chain reports (124/150, 82.67\% overall precision). SCAN-MCP achieved the highest observed precision among the evaluated analyzers on this subset. Sink-only controls reproduced the asserted effect in 0 of the 19 confirmed tool-chain cases under the tested conditions. Removing slicing or proposal filters reduced precision in the ablation study.
These results show that securing the MCP ecosystem requires analyzing tool compositions, not only individual tool implementations.
\end{abstract}

\section{Introduction}
% no \IEEEPARstart
The transformation of Large Language Models (LLMs) from text generators into autonomous agents is driven by their ability to use external tools. This agent-based architecture requires standardized middleware to coordinate tool invocation, context retrieval, and multi-step workflows. The Model Context Protocol (MCP) is rapidly emerging to fill this role, effectively serving as the ``USB-C for AI''~\cite{mcp}. Its adoption is accelerating across major agentic frameworks, including those from OpenAI, Anthropic, Microsoft AutoGen, and LangChain~\cite{ref-openai-api, ref-anthropic-tools, ref-autogen-api, ref-langchain-api}, while public registries already list thousands of independently built MCP servers~\cite{ref-community-registry}. As MCP becomes common infrastructure for agentic AI, it has also become an active security research target, with recent work examining single-tool flaws and server-level risks such as overbroad file access, unsafe command execution, and inadequate input validation~\cite{li2025toward,hasan2025model,microsoft2025}. However, composition across tools remains less understood.

\vspace{2pt}\noindent \textbf{Emerging composition risks.} This composition gap is consequential. In practice, an LLM or user often invokes multiple tools sequentially, passing intermediate artifacts through files, shared state, parameters, or conversational context~\cite{wang2025mcp}. Under this execution model, a tool that appears benign in isolation can become one step in an exploit chain when another tool consumes its output or relies on the state it creates. These cross-tool edges are selected at run time by an LLM or user and are not represented as source-level calls, so a conventional call graph cannot recover the chain even when each tool's local behavior is visible. Our study shows that such tool chaining is common, and that the resulting vulnerabilities are often consequences of unsafe composition among otherwise ordinary tools rather than isolated coding errors.

\begin{figure}[t]
  \centering
  \includegraphics[width=0.82\linewidth]{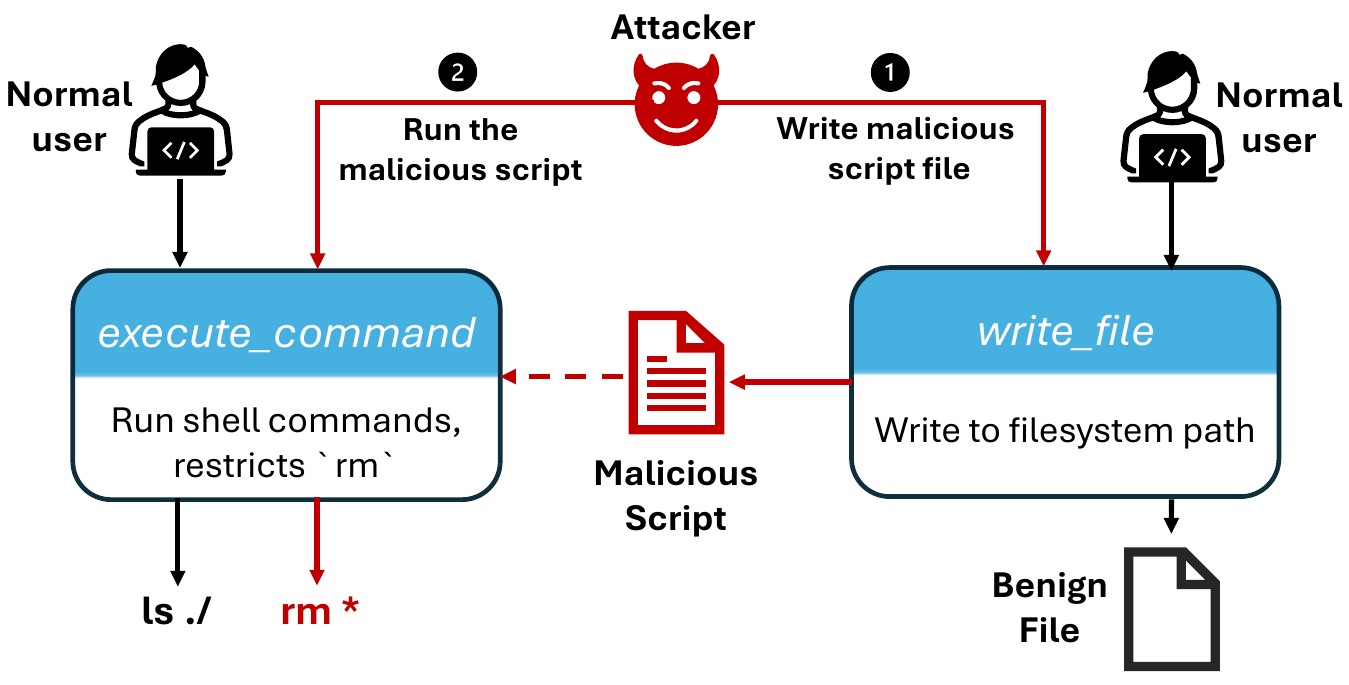}
  \caption{Tool-chain bypass of command restrictions via an attacker-written script}
  \label{fig:motivating}
\end{figure}

\vspace{2pt}\noindent \textbf{Motivating example.} Consider a terminal-style MCP server with two tools. The first, \texttt{execute\_command}, runs shell commands but blacklists dangerous command strings such as \texttt{rm}; it checks the invocation text, not the contents of files that invocation executes. The second, \texttt{write\_file}, writes user-supplied content to a filesystem path. In isolation, \texttt{write\_file} only creates files, and \texttt{execute\_command} rejects direct unsafe commands. In composition, however, an attacker can use \texttt{write\_file} to create a script containing a forbidden operation, then execute that script through \texttt{execute\_command}. The command string no longer contains the forbidden token, so the blacklist is bypassed and actions such as file deletion become possible. 
The example illustrates how state established by one tool can affect the security outcome of a subsequent tool invocation (Fig.~\ref{fig:motivating}).

\vspace{2pt}\noindent \textbf{Challenges and our solution.} A natural mitigation is to apply program analysis or rule-based scanning to MCP tools~\cite{hasan2025model}, but this approach faces three challenges. First, MCP servers are implemented in diverse languages and use many registration idioms, including decorators, factories, configuration loaders, and structured exports, which obscures tool entry points and makes fixed rule sets brittle~\cite{li2022polycruise,lee2020broadening}. Second, tool composition occurs at run time: invocation order is chosen by an LLM or user, and intermediate artifacts often pass through files, shared state, or conversational context rather than explicit source-level call edges. Third, the ecosystem contains tens of thousands of tools, making manual review or per-server customization infeasible. To handle heterogeneous registration, SCAN-MCP identifies tool registrations, maps tool names to concrete handlers, and extracts Tree-sitter-based code slices. To recover run-time composition, it performs semantic source and sink inference and constructs inter-tool dependency graphs. To screen proposals at ecosystem scale, SCAN-MCP combines source-level checks, structural and type constraints, and pattern filters. Selected reports undergo manual sandbox confirmation.

\vspace{2pt}\noindent \textbf{Findings.}
We ran SCAN-MCP on a corpus of 10,121 MCP servers with 90,200 discovered tools. Sources represented in the final corpus include MCP.so, Glama, Smithery, Pulse, and GitHub (Table~\ref{tab:rq2-platform}). Stage 3 produced 27,123 reported findings, including 8,422 tool-chain reports (31.05\%).
These reports cover issues such as command execution, overbroad file access, and cross-tool data-flow abuse. This fraction describes the distribution of SCAN-MCP's reports rather than the prevalence of confirmed vulnerabilities in the ecosystem. 
On a language- and marketplace-stratified subset of 100 servers drawn from this corpus, we manually evaluated 150 Stage 3 reports and confirmed 124 (82.67\% precision). The full-corpus total and the sample total both count Stage 3 reports with proof-of-concept sketches and asserted observable effects.
The results show that tool-chain findings are a recurring pattern in SCAN-MCP's reports: they arise from unguarded dangerous APIs, unsafe parameter propagation, and per-tool checks that can be bypassed by co-located tools.

\vspace{2pt}\noindent\textbf{Contributions}. The contributions of this paper are summarized as follows:

\noindent\textbullet\ \textit{A graph-based formulation of MCP composition analysis.}
We formulate the search for candidate tool chains as bounded reachability in a typed inter-tool graph. Nodes represent exposed tools of a single MCP server, and edges encode proposed dataflow or enablement relations. Structural and type checks prune candidate paths before report generation.

\noindent\textbullet\ \textit{Tool-chain-aware analysis framework.} We design SCAN-MCP, a language-agnostic analysis framework that recovers tool-handler bindings across heterogeneous registration idioms, lifts each tool into a normalized source and sink representation, and reasons over an inter-tool graph to surface composition-dependent flows. The pipeline combines source-level checks, structural and type gates, and separate sandbox confirmation for evaluated reports.

\noindent\textbullet\ \textit{Ecosystem-scale empirical findings.} We conduct an ecosystem-scale analysis of 10{,}121 MCP servers and 90{,}200 tools, producing 27{,}123 reported findings (defined in Section~\ref{sec:scan-mcp}) and a manually adjudicated sample of 150 reports with 124 sandbox-confirmed findings. The study provides empirical evidence that securing MCP deployments requires analyzing tool chains rather than only isolated tool implementations, and we release our analysis framework to support replication~\cite{scanmcp_artifact}.

\vspace{2pt}\noindent\textbf{Artifact availability, ethics, and responsible disclosure.}
To support verification and replication, we provide an artifact package at \url{https://github.com/mmzha2013/scan-mcp}~\cite{scanmcp_artifact}. The artifact contains the SCAN-MCP implementation, prompt templates, and scripts for reproducing the analysis pipeline, and three sandbox demonstration videos corresponding to the case studies in Section~\ref{sec:evaluation-c}: arbitrary command execution, the write-and-execute tool-chain bypass, and the browse-click-upload file-disclosure chain. We analyze local clones of public repositories and do not probe live deployments. We manually validate selected reports only in isolated sandboxes. Before submitting code slices to LLMs, we screen for hard-coded secrets using GitLeaks~\cite{gitleaks} to reduce the risk of exposing sensitive data. We responsibly report confirmed findings to affected maintainers and MCP marketplaces, including MCP.so~\cite{mcp-so}, Glama~\cite{glama}, and Smithery~\cite{smithery}. One confirmed finding involving a widely used browser-automation MCP server was initially deemed out of scope, yet the same unsafe path-handling behavior was later patched, providing independent evidence that such composition flaws are practically relevant.
\section{Background}

\subsection{Model Context Protocol Overview}
The \textit{Model Context Protocol} (MCP) standardizes how LLM applications and agents connect to external capabilities. An MCP server exposes callable \textit{tools}, each defined by a schema (name, parameters, and description) and implemented by an executable \textit{handler}; clients such as LLM runtimes, IDEs, and orchestration agents discover these schemas and invoke tools through JSON-RPC over HTTP, WebSocket, or stdio~\cite{mcp-client,mcp-server,mcp-tool}. A typical deployment pairs a host application, such as Codex or Claude Desktop, with one or more MCP servers~\cite{visualstudio-code,claude-desktop}. This design decouples model reasoning from system capabilities, but it also enables \textit{tool chaining}, where the output, side effect, or intermediate artifact of one tool is later consumed by another within a workflow orchestrated by an LLM or user.

\subsection{MCP Security and Limitations}
MCP standardizes tool invocation but leaves most security enforcement to individual servers, including input validation, capability restriction, authentication, and inter-tool privilege boundaries~\cite{mcp-server}. Its OAuth~2.1 authorization model is optional and, by public accounts, complex and inconsistently adopted, so many deployments only partially implement authentication and leave tool-chain interactions weakly governed~\cite{mcp-auth,mcp-limitation1,mcp-limitation3}. Existing analyses have primarily examined single-tool risks, such as unsafe command execution and arbitrary file writes~\cite{ferrag2025prompt,narajala2025enterprise,song2025beyond}. For our setting, the key limitation is that a server may validate each tool locally while still allowing unsafe behavior when tools exchange files, parameters, shared state, or conversational context. This limitation motivates an analysis that considers both individual tool handlers and inter-tool interactions.

\section{Threat Model}
\label{sec:threatmodel}
We consider adversaries that exploit \emph{legitimate} MCP server tools to compromise host resources or mislead downstream LLM clients through untrusted tool outputs. \textbf{Assets} include host state (files, secrets, configurations, and processes), tool outputs and provenance metadata that an LLM client may consume as context, LLM runtime state (prompts and context buffers), and user-controlled inputs such as files, URLs, and tokens. \textbf{Trust assumptions:} the server is not malicious and executes its published tools as implemented, but tool outputs are not inherently attested and may be treated as trusted when provenance checks, schema validation, or capability boundaries are absent. \textbf{Attack scenarios:} first, a single tool can be exploited when crafted parameters trigger unsafe functionality such as arbitrary file access, command execution, or insecure deserialization; 
second, a security impact may depend on an artifact or state established by an earlier tool in the sequence.
\textbf{Goals:} the adversary seeks to violate confidentiality by extracting secrets or user data, integrity by modifying files, forging tool outputs, or misleading downstream LLM decisions, or availability by causing resource exhaustion or unsafe execution. \textbf{Capabilities and scope:} the adversary is a remote, non-privileged user who can invoke tools through normal prompts or APIs and can read public code. The adversary cannot modify server code, deploy binaries, or obtain shell access except by invoking an exposed tool whose implemented functionality reaches shell execution; physical compromise, insider access, and social engineering are out of scope.
\section{Motivation and Design Challenges}
\label{sec:challenges}

\subsection{Why MCP Composition Is Not Classic Interprocedural Analysis}

MCP composition differs from conventional interprocedural analysis because the relevant edges are not materialized as source-level calls, returns, or explicit parameters. In classic software, data-flow edges between functions appear in the program text, so interprocedural analyzers such as FlowDroid~\cite{arzt2014flowdroid} or cross-language frameworks such as PolyCruise~\cite{li2022polycruise} can track information flow across procedure boundaries. In MCP, tools are independently registered entry points. An LLM or user chooses the invocation order at run time, and intermediate artifacts may move through files, paths, environment variables, browser state, shared databases, or conversational context rather than explicit function parameters. A tool that is locally benign can therefore establish state or produce an artifact that a later tool treats as trusted input.

Consider the motivating example: even with the complete source code of both \texttt{write\_file} and \texttt{execute\_command}, a conventional call-graph analysis finds no edge between them, because neither function calls the other. The exploitable relation emerges only when a user or LLM invokes them in sequence, a decision absent from the code. Precise handler recovery and handler-level slicing are therefore necessary but insufficient: they expose each tool's local behavior, but the composition edge exists in no source file and must be reconstructed separately over an inter-tool representation.

\subsection{Design Challenges}

Discovering composition-dependent MCP vulnerabilities at ecosystem scale imposes three design constraints.

\vspace{2pt}\noindent\textbf{C1: Heterogeneous implementations.}
MCP servers span Python, TypeScript/JavaScript, C\#, PHP, and other languages, echoing the heterogeneity faced by multilingual program analysis~\cite{li2022polycruise,lee2020broadening}. Beyond language diversity, MCP tools may be registered through decorators, factory exports, configuration files, framework wrappers, or SDK abstractions. This heterogeneity constrains the frontend: before any cross-tool reasoning, an analyzer must resolve exposed tools and handlers across diverse registration idioms. A rule set tied to one language or framework can therefore fail at the entry point rather than at the flow-tracking stage.

\vspace{2pt}\noindent\textbf{C2: Semantic cross-tool dependencies.}
MCP chains are usually not encoded as function calls. An LLM or user selects the tool order at run time, and data can flow through files, parameters, shared state, third-party libraries, session state, or conversational history~\cite{li2025dissonances}. This requires a representation that models dataflow and enablement relations among independently exposed tools, not only local operations inside each handler.

\vspace{2pt}\noindent\textbf{C3: Scale, precision, and validation cost.}
Large MCP ecosystems force composition-aware analysis to control both search cost and false-positive rate. The number of candidate tool pairs grows quadratically with the number of tools per server, and manually validating every candidate chain is infeasible because each finding may require code inspection, environment setup, and exploit confirmation. An effective analyzer must therefore prune aggressively, ground candidates in source evidence, and distinguish reported findings from sandbox-confirmed findings.

\vspace{2pt}\noindent\textbf{Our approach.}
SCAN-MCP addresses C1 with an LLM-assisted frontend that maps tool schemas to handlers and uses static slicing to recover normalized per-tool behavior. It addresses C2 by constructing an inter-tool dependency graph from inferred sources, sinks, artifacts, and shared state. It addresses C3 by applying stage-specific checks to model proposals and recording distinct assurance levels for candidates, reports, and sandbox-confirmed findings.
\section{SCAN-MCP}
\label{sec:scan-mcp}
\begin{figure*}[t]
  \centering
  \includegraphics[width=0.72\linewidth]{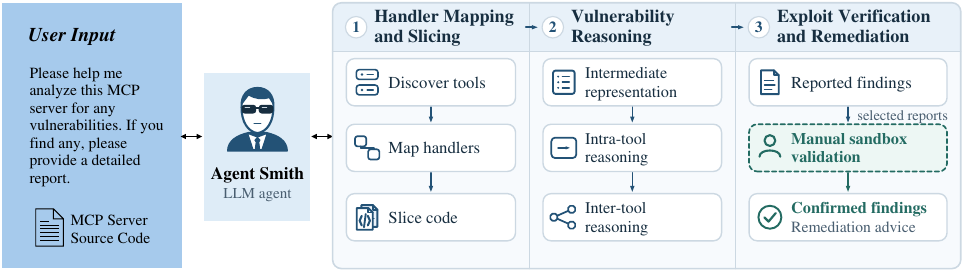}
  \caption{Architecture of SCAN-MCP. The framework maps MCP tool schemas to handlers, extracts per-tool slices, proposes intra-tool and inter-tool flows, generates candidate PoCs, and supports manual sandbox validation of selected reports.}
  \label{fig:mcp-arch}
\end{figure*}

To address the heterogeneous, composition-dependent, and ecosystem-scale challenges in Section~\ref{sec:challenges}, we design \textbf{SCAN-MCP}, a three-stage framework for analyzing MCP servers (Fig.~\ref{fig:mcp-arch}). SCAN-MCP normalizes repositories into per-tool code slices, uses LLM-assisted reasoning with deterministic structural and type-based gates to produce intra-tool and inter-tool candidates, and supports manual sandbox validation of selected PoCs when execution is feasible. 
The LLM proposes mappings, summaries, flows, and proof-of-concept sketches. The pipeline checks handler bindings against source code and applies structural and type checks to candidate chains. Prompts request code citations for semantic proposals. Analysts assess selected reports separately in an isolated sandbox.

\noindent\textbf{Levels of assurance.}
SCAN-MCP reports three output levels. We use \emph{vulnerability candidate} as an umbrella term for an unconfirmed SCAN-MCP output, and distinguish the following levels when precision matters.
(i) \emph{Candidate}: a flow or chain that survives the deterministic structural and type-based gates, indicating plausible source, sink, and graph structure but not runtime exploitability.
(ii) \emph{Reported finding}: a candidate with an LLM-generated PoC sketch and asserted observable effect, such as a sentinel file, spawned process, modified state, or outbound probe. Full-corpus counts are reported at this level unless explicitly labeled as confirmed.
(iii) \emph{Confirmed finding}: a reported finding whose asserted effect is reproduced by a human analyst in an isolated sandbox. The 100-server precision measurement is computed at this level; reported findings are never counted as confirmed without sandbox reproduction.

Fig.~\ref{fig:mcp-arch} shows the overall architecture. SCAN-MCP ingests an MCP server repository, recovers tool registrations and handler bindings, extracts source-level slices for each resolved tool, constructs semantic summaries and an inter-tool dependency graph, and produces candidate PoCs for analyst validation. Each stage produces structured outputs so intermediate results can be inspected, cached, and reused across servers.

\subsection{Handler Mapping and Slicing}

\vspace{2pt}\noindent\textbf{Goal and inputs.} The objective of Stage~1 is to transform a repository that hosts an MCP server into a structured collection of \emph{tool slices}. The input is the source code of an MCP server. The output is a set of records, each binding a schema-level tool name to a concrete handler location and a source-level slice that contains the handler body and any transitively invoked local functions recovered by the slicer. Later stages operate on these normalized slices.

\vspace{2pt}\noindent\textbf{Why an agent is needed.} Challenge~C1 shows that MCP tools are registered through diverse language and SDK idioms, and later stages depend on resolving the correct handler for each tool. A single static locator is brittle because tool registration may be split across decorators, factory exports, configuration files, dispatch branches, and framework wrappers. Prior work shows that large language models are effective at semantic pattern matching and binding inference~\cite{pandey2025design,jelodar2025large}, so SCAN-MCP uses an agent to coordinate lightweight static discovery, LLM-assisted binding proposals, and local source checks across files. The LLM helps identify likely tool-handler bindings from code context, but a binding is accepted only after the reported signature, callback, or dispatch branch is checked against the original source.

\vspace{2pt}\noindent\textbf{Tool discovery and handler mapping.}
SCAN-MCP first identifies candidate MCP tools by scanning source files and configuration files for common registration patterns, including SDK calls, decorator-style registrations, exported tool lists, schema objects, and dispatcher branches. For each matched registration context, it extracts the schema-level tool name, the file location, and nearby binding cues such as a callback expression, handler identifier, parameter schema, description field, or dispatch key. SCAN-MCP then treats each extracted tool name $t$ as an anchor and searches the registration file and its first-party import closure for candidate handlers. Third-party and standard-library modules are treated as opaque and are not expanded. Each candidate handler or dispatch branch is represented as a \emph{binding proposal} for tool $t$ through the registration anchor and the LLM-assisted binding step. This proposal step defines the candidate set rather than the final label: a missed proposal can reduce recall, which we measure in Table~\ref{tab:rq1-stage-metrics}, but no proposal is accepted without source checks. Given a proposed handler $h$ for $t$, SCAN-MCP accepts the binding only if three deterministic checks hold: (i)~$h$ is located in the registration file or in a reachable first-party module; (ii)~the surrounding code links $h$ to $t$'s schema, name, description, or dispatch path; and (iii)~the reported handler location is re-checked against the original source, and the extracted signature, callback, or dispatch branch appears at that location or matches a local symbol resolved from that location. The LLM rationale and confidence are recorded for auditing, but are not acceptance criteria. The search terminates when a proposed handler passes the three checks, or when the bounded import closure is exhausted. If no proposed handler passes all three checks, the tool is marked unresolved and excluded from downstream vulnerability reasoning.

\vspace{2pt}\noindent\textbf{Static slicing.} Given a checked handler, SCAN-MCP treats it as the entry point for static slicing. It uses Tree-sitter to build a local call graph and follows direct function calls, method invocations on \texttt{this} or \texttt{self}, and imports to sibling files within the same module or package. SCAN-MCP represents calls into third-party or standard-library code as annotated call sites containing the callee identifier, module path, and argument expressions. Section~\ref{sec:evaluation-B} reports the call-depth audit and the supplementary static analysis. The resulting slice gives Stage~2 a compact, auditable view of each tool's repository-local behavior.

\subsection{Vulnerability Reasoning}
Stage 2 uses normalized tool slices to propose candidate flows within individual tools and candidate chains spanning multiple tools.
It first converts each slice into a language-agnostic intermediate representation, then proposes source-to-sink flows within individual tools, and finally constructs cross-tool edges among tool pairs and enumerates candidate chains over them. In all three steps, LLM outputs are treated as hypotheses that are filtered by source-grounding requirements and deterministic, type-based gates before they are retained as candidates.

\vspace{2pt}\noindent\textbf{Step I: Intermediate representation.} For each tool slice, SCAN-MCP builds an intermediate representation with three fields: \texttt{description}, \texttt{sources}, and \texttt{sinks}. The description is a short summary of the tool's purpose. Each source or sink entry contains a \texttt{type} drawn from a fixed taxonomy, such as untrusted user input, file read, file write, network request, or command execution, 
and a \texttt{snippet\_evidence} field for a code citation that the prompt asks the LLM to draw from the tool slice. This representation captures security-relevant behavior at the semantic level while recording the evidence cited for the proposed sources and sinks.

\vspace{2pt}\noindent\emph{Design rationale.} LLM guidance is used in Step~I because developer-supplied descriptions are optional and often omit security-relevant behavior; documentation for an MCP SDK notes that tools may have a confusing or absent description~\cite{gofastmcp_tool_transformation}. A normalized summary and source or sink proposal gives later stages a consistent representation across languages and SDKs. Because these proposals may be incomplete or unsupported, SCAN-MCP applies the source-grounding requirements and type-based gates described in Step III before retaining them as candidates.

\vspace{2pt}\noindent\emph{Source-grounding requirements and pattern filters.}
SCAN-MCP requires each proposed source or sink to cite code evidence from the slice, enforced primarily through the LLM prompt rather than a complete programmatic substring proof. Automated checks are narrower: language-specific keyword and regex patterns identify known source and sink operations such as subprocess calls, file operations, network requests, and dynamic evaluation, while deterministic scans reject malformed or forbidden PoC prompts. These are evidence filters, not a full type-system or schema-compatibility proof.

\vspace{2pt}\noindent\textbf{Step II: Intra-tool reasoning.}
For each retained slice representation, SCAN-MCP asks the LLM to propose candidate source-to-sink flows within the tool and to provide a rationale for each flow. A flow is retained as a candidate only if its endpoints correspond to retained source or sink entries, its rationale cites source evidence from the slice, and the sink represents an operation that can affect confidentiality, integrity, or availability. For retained flows, SCAN-MCP records the candidate flow together with its source and sink evidence. The flow remains a candidate until Stage~3 synthesizes a PoC consisting of tool parameters, execution steps, and an asserted observable effect, turning it into a reported finding that sandbox validation may later confirm.

\vspace{2pt}\noindent\textbf{Step III: Inter-tool reasoning.} To capture composition-dependent candidates, SCAN-MCP builds a directed graph whose nodes are tools and whose edges denote potential cross-tool dataflow or enablement relations. For every ordered pair of tools $(A,B)$ in a server, SCAN-MCP asks the LLM whether $A$'s output or side effect could feed $B$ through an explicit or implicit shared artifact, such as a written file, a temporary directory, an environment variable, a database record, or a shared state key, or whether $A$ could satisfy a precondition required for $B$'s later execution. Each ordered pair is queried in a fixed producer$\rightarrow$consumer direction, so the relation direction is determined by the query structure rather than freely chosen by the model. The LLM returns two boolean relation flags, dataflow and enablement, each with a confidence score and a free-text rationale. An edge is retained under Gate~1 (the relation-flag criterion): at least one of the two flags (\textit{dataflow} or \textit{enablement}) must be set. Throughout, a composition is considered \emph{relevant} exactly when Gate~1 holds, so ``relevant'' denotes this checkable flag condition rather than a free-form model judgment. The rationale and confidence are recorded only for auditing: they are not used as accept or reject criteria, and SCAN-MCP performs no programmatic rationale-to-slice match at this step. 
Pair-level proposals use tool names and functional summaries. Chain enumeration applies type and graph constraints. Subsequent chain analysis receives the available code slices for the tools in each chain. Sandbox observations support the individual cases tested.

\vspace{2pt}\noindent\emph{Chain enumeration and deterministic gates.} After edge acceptance, SCAN-MCP enumerates candidate chains over the directed inter-tool graph. A possible entry node must contain a retained source type from an untrusted-input allow list (Gate~2), and a possible endpoint must contain a retained sink type from a server-impacting or network-exposing allow list (Gate~3). 
In the accompanying artifact, SCAN-MCP keeps reachable simple paths (Gate~4) that traverse at least one dataflow edge (Gate~5) and contain at most three inter-tool edges (Gate~6), corresponding to at most four tools. Section~\ref{sec:evaluation-c} reports chain size as the number of tools in the reported sequence. These gates are implemented as deterministic code over the type taxonomy and graph structure. They bound the search and remove structurally implausible chains through type and connectivity checks. Sandbox validation assesses runtime exploitability in the cases tested.

\vspace{2pt}\noindent\emph{Reliability and reproducibility.}
All LLM calls use temperature $=0$ and fixed prompt templates, and each records the provider, model identifier, and prompt version for auditing. Because hosted models do not guarantee bit-for-bit reproducibility, SCAN-MCP does not rely on decoding determinism: final confirmation labels never rest on the LLM response alone, but on source-grounding requirements, deterministic type-based gates, and manual sandbox confirmation.

\subsection{Exploit Verification and Remediation}
Stage 3 converts retained candidates into reported findings. For each candidate, the LLM proposes an invocation sequence, concrete inputs, and an asserted observable effect such as a sentinel file, spawned process, modified state, or outbound probe. This produces a reported finding; it is not confirmation. Because MCP servers often require deployment-specific setup, confirmation is performed manually in an isolated VMware VM. The analyst reconstructs the required environment from declared dependencies and documentation, invokes the tools through the native MCP interface, and checks whether the asserted effect occurs. A reported finding is counted as a \emph{confirmed finding} only when the observed effect matches the PoC assertion; otherwise it remains unconfirmed. For confirmed findings, SCAN-MCP reports the PoC, supporting evidence, and remediation suggestions, but does not apply patches automatically.

\section{Evaluation and Findings}
\label{sec:evaluation}
This section evaluates SCAN-MCP along three dimensions: detection effectiveness, large-scale security findings, and scalability.

\subsection{Research Questions and Setup}
\noindent\textbf{Research Questions.} Our evaluation is guided by three research questions: 
\textbf{RQ1:} How effective is SCAN-MCP in finding vulnerabilities within and across MCP tools compared with existing analyzers? 
\textbf{RQ2:} What vulnerability patterns emerge from large-scale analysis of real-world MCP servers? 
\textbf{RQ3:} How well does SCAN-MCP scale to large, heterogeneous MCP deployments?

\vspace{3pt}\noindent\textbf{Data and Corpus.} 
Between April and July 2025, we collected 17,972 candidate project records from the GitHub API and four MCP marketplaces (MCP.so, Glama, Smithery, and Pulse). Initial screening for valid, reachable links and a verifiable
MCP server startup interface retained 14,273 MCP server entries and excluded 3,699 candidates. We then removed 3,450 entries without tool capability, leaving 10,823 entries. Server-level deduplication removed 702 duplicate entries, yielding a final corpus of 10,121 MCP servers with 90,200 discovered tools. Table~\ref{tab:rq2-platform} groups tool-bearing repository paths by the sources recorded in the collection metadata. A path can appear under more than one source. MCP.so has the largest membership, followed by Pulse and Glama.
Language labels are not mutually exclusive because a server may contain multiple implementation languages; \textit{TypeScript} appears in \textit{5,771} servers (\textit{57.02}\%), followed by \textit{Python} (\textit{43.67}\%) and \textit{JavaScript} (\textit{6.43}\%). This distribution is consistent with rapid adoption in scripting-oriented and web-oriented ecosystems, where low barriers to tool creation may also increase the risk of insecure implementations and misconfigurations.

Of the 90,200 discovered tools, 9,654 (10.70\%) were marked unresolved and excluded before slicing, and 49 (0.054\%) had no mapping entry. A separate manual audit found handler or dispatch source evidence for all 754 tools in the 100-server sample. This audit measured source visibility, while Table~\ref{tab:rq1-stage-metrics} reports component performance.

\vspace{3pt}\noindent\textbf{Baselines and Metrics.}
We compare SCAN-MCP with four detectors: Semgrep~\cite{semgrep}, a rule-based static analyzer; AI-Infra-Guard (A.I.G)~\cite{aig-home,aig}, an LLM-based analyzer from Tencent; mcp-scanner~\cite{cisco-mcp-scanner}, a Cisco MCP scanner run in source-level behavioral mode; and Pure-LLM, which prompts the same reasoning model over raw repositories without SCAN-MCP's handler recovery, slicing, source-grounding, filters, or inter-tool graph. Direct baselines must run on repository snapshots and emit per-tool findings; mcp-scanner qualifies because its behavioral analyzer emits source-level per-tool verdicts. We exclude snyk-agent-scan~\cite{snyk-agent-scan} and mcp-shield~\cite{mcp-shield} because they target deployed configurations, permissions, or runtime posture, and discuss them as related work. We use the three assurance levels defined in Section~\ref{sec:scan-mcp}; for baseline comparison, the same reported/confirmed criteria are applied uniformly. 
For $R$ reports and $C$ confirmations, precision is $C/R$ and the false discovery rate (FDR) is operationalized as $(R-C)/R$ under our sandbox protocol. For SCAN-MCP, 124 of 150 reports were confirmed (82.67\%) and 26 remained unconfirmed (17.33\%). Reports blocked by unmet runtime prerequisites are included in the unconfirmed count.
Full-corpus results report SCAN-MCP's reported findings unless explicitly labeled as confirmed.

\vspace{2pt}\noindent\textbf{Baseline Protocol.}
All detectors are evaluated on the same 100-server stratified sample and repository snapshots. We run Semgrep v1.86.0 and AI-Infra-Guard v2.6 on the same repository inputs, and run Pure-LLM with the same primary reasoning model used by SCAN-MCP for vulnerability reasoning. We run mcp-scanner v4.7.4 in its behavioral mode—its source-level analyzer, combining interprocedural cross-file dataflow tracking with LLM-based alignment checking—backed by the same class of reasoning model. For comparability, we normalize all outputs into a common finding schema with server, tool, vulnerability category, available supporting evidence, and single-tool/tool-chain label. We then apply the same deduplication and sandbox-validation criteria to SCAN-MCP and all baselines before computing confirmed-over-reported counts, FDR, and precision.

\vspace{3pt}\noindent\textbf{Implementation Details.} SCAN-MCP is implemented in Python. Tool discovery and handler mapping use registration-pattern matching, LLM-assisted binding proposals, and local signature or dispatcher checks; static slicing uses \textit{Tree-sitter v0.25.4} across ten programming languages to build local call graphs and normalize code slices. We use \textit{Qwen3-coder} for tool-handler mapping, \textit{Grok-4-fast} as the primary reasoning model for vulnerability identification and PoC synthesis, and \textit{gpt-oss-120b} for lightweight consistency checks. Prompts use fixed templates with temperature $=0$ and caching to reduce variance and cost. All experiments ran on an Ubuntu server with an \textit{Intel(R) Xeon(R) Gold 5222 CPU @ 3.80~GHz} and \textit{125~GB} RAM. PoC confirmation was performed manually inside isolated VMware virtual machines to contain side effects.

\vspace{3pt}\noindent\textbf{Sampling for Comparative Evaluation.}
For controlled comparison with baselines, we drew a stratified random sample of 100 MCP servers from the corpus. When a server appeared in multiple collection sources, we assigned it to the source in which it first appeared in our collection records. We allocated sample slots across sources in proportion to the number of servers assigned to each source, then distributed each source's slots across implementation languages according to their proportions within that source. We used a fixed random seed of 2025 to draw the sample.
On this sample, SCAN-MCP emitted 150 reported findings, of which 124 were confirmed by manual validation, yielding 82.67\% precision. 
The 95\% confidence interval for report-level precision [76.67\%, 88.67\%] was obtained by bootstrapping the 150 reports with 10,000 resamples. We used this 100-server subset for baseline comparisons, component evaluation, and ablation. RQ2 describes the distribution of Stage 3 reports across the full corpus.

\subsection{RQ1: Detection Effectiveness}
\label{sec:evaluation-B}
\noindent\textbf{Objective.} RQ1 evaluates SCAN-MCP's ability to detect single-tool and tool-chain vulnerabilities and quantifies the contribution of its components. We combine baselines, component-level measurements, ablation, and error analysis.

\vspace{2pt}\noindent\textbf{Baseline Comparison.} Table~\ref{tab:rq1-effectiveness} compares SCAN-MCP with Semgrep, A.I.G, mcp-scanner, and Pure-LLM on the same 100-server stratified sample. The table reports confirmed-over-reported tool-chain vulnerabilities (TV), single-tool vulnerabilities (SV), FDR, and precision.

\begin{table}[t]
\centering
\caption{Baseline comparison on a 100-server stratified sample. \#TV and \#SV report confirmed/total reported findings.}
\label{tab:rq1-effectiveness}
\small
\begin{tabular}{lcccc}
\toprule
\textbf{Method} & \textbf{\# TV} & \textbf{\# SV} & \textbf{FDR (\%)} & \textbf{Precision (\%)} \\
\midrule
Semgrep              & 0/0            & 27/141           & 80.85 & 19.15 \\
A.I.G                & 0/0            & 70/191           & 63.35 & 36.65 \\
mcp-scanner                & 0/0            & 0/2              & 100.0 & 0.0 \\
Pure-LLM             & 6/8            & 58/118           & 49.21 & 50.79 \\
\textbf{SCAN-MCP}    & \textbf{19/20} & \textbf{105/130} & \textbf{17.33} & \textbf{82.67} \\
\bottomrule
\end{tabular}
\end{table}

SCAN-MCP achieves the highest precision and the largest number of confirmed findings. 
For confirmed single-tool findings, Semgrep, A.I.G, and Pure-LLM shared 22, 27, and 14 findings with SCAN-MCP, respectively. In the same order, the baseline-only counts were 5, 43, and 44, and the SCAN-MCP-only counts were 83, 78, and 91. SCAN-MCP recovered all six tool-chain findings confirmed by Pure-LLM and confirmed 13 additional tool-chain findings in that comparison. The tool-chain results for Semgrep, A.I.G, and mcp-scanner reflect the tool-local scope of their evaluated configurations. In the evaluated configuration, mcp-scanner checked whether each tool's implementation matched its description. It produced two reports, both of which we classified as false positives in the error analysis.

The 19 confirmed tool-chain findings occurred across nine of the 100 sampled servers. To assess composition dependence, we compared the complete sequence with the final tool alone, using the same clean initial state and attacker privileges for each paired comparison. The sink-only controls included schema-valid, repeated, and encoded inputs. In the original sandbox validation, the complete sequences reproduced the asserted effects for these 19 confirmed SCAN-MCP tool-chain findings. In follow-up sink-only controls, the asserted effect was reproduced in 0 of 19 cases under the tested conditions. This supports composition dependence for the evaluated cases. 

\vspace{2pt}\noindent\textbf{Error Analysis.}
Residual errors differ by detector. A.I.G over-flags benign behavior, such as treating ordinary error messages as disclosure; Semgrep flags dangerous APIs that are unreachable or constrained in context; and Pure-LLM proposes cross-tool chains without concrete shared artifacts or executable preconditions. mcp-scanner's two reports were false positives caused by over-interpreting benign API-key use and by mapping an anonymous code fragment to a non-exposed tool. 
Unconfirmed SCAN-MCP reports included cases blocked by unmet runtime prerequisites, such as unavailable authentication or storage services.

\vspace{2pt}\noindent\textbf{Component-wise Evaluation.} We measured individual stages against manually inspected ground truth. 
The component evaluation included all 754 tools exposed by the sampled servers. Two researchers independently labeled tool presence, handler bindings, reachable code, source/sink entries, and cross-tool relations. They resolved disagreements through discussion and used the agreed labels as the reference for the evaluated stages. For edge inference, we manually reviewed all retained inter-tool relations in the 100-server sample. The retained predictions included dataflow-only relations, enablement-only relations, and relations marked with both labels. They came from 59 servers exposing 2 to 32 tools each.
Structural reconstruction stages report precision and recall; semantic stages are precision-oriented because exhaustively enumerating all latent sources, sinks, edges, and exploitable chains in in-the-wild repositories is not feasible.

\begin{table}[t]
\centering
\caption{Per-stage effectiveness of SCAN-MCP. Semantic stages are evaluated with precision-oriented metrics.}
\label{tab:rq1-stage-metrics}
\small
\begin{tabular*}{\columnwidth}{@{\extracolsep{\fill}}ccc}
\toprule
\textbf{Stage} & \textbf{Precision (\%)} & \textbf{Recall (\%)} \\
\midrule
Tool extraction         & 100.00 & 100.00 \\
Handler mapping         & 100.00 & 99.60  \\
Tool slicing            & 100.00 & 100.00 \\
Source-and-sink labeling & 96.00 & -- \\
Edge inference          & 98.39  & -- \\
Sandbox confirmation    & 82.67  & -- \\
\bottomrule
\end{tabular*}
\end{table}

On the manually inspected outputs, source/sink labeling achieved 96.0\% precision and edge inference achieved about 98.39\% precision. We compared the text cited in 2,792 source/sink annotations with the available code. The cited text matched verbatim in 2,418 annotations and after normalization in another 200. In a targeted manual audit of 38 distinct nonliteral citations, we found 35 grounded citations and three that pointed to unrelated handlers in oversized slices. Beyond grounding the individual annotations, assessing a proposed flow
requires evidence that the source can reach the sink. The 124/150 sandbox confirmation fraction is an end-to-end measure on reported findings.

To examine slicing depth, we recursively extended 668 saved tool trees from the 754-tool sample, preserving their handler roots and first-level edges. Depths of 0 through 5 occurred in 241, 235, 33, 132, 20, and 7 of the resulting trees, respectively. The median depth was 1 and the 95th percentile was 3. Depth caps of 10, 20, and 40 yielded identical trees.

\vspace{2pt}\noindent\textbf{Ablation Study.} Table~\ref{tab:rq1-ablation} reports ablations with the model and 100-server corpus fixed. V1 removes static slicing, V2 removes source-grounding and pattern filters, V3 removes inter-tool reasoning, and V4 is an LLM-only reference baseline over raw repositories.
V1 without slicing produced 215 reports and confirmed 46, all single-tool findings (21.40\% overall precision). V2 produced 519 reports, 369 more than Full, while precision decreased from 82.67\% to 32.37\%. All 124 findings confirmed by Full were also confirmed by V2. V2 confirmed 44 additional findings (36 SVs and 8 TVs). Full therefore covered 124 of V2's 168 confirmed findings (73.8\%). V3 without inter-tool reasoning retained 105 confirmed SVs among 130 SV reports and emitted zero TV reports. The V4 LLM-only reference confirmed six tool-chain findings, all shared with Full.

\begin{table}[t]
\centering\footnotesize
\setlength{\tabcolsep}{3pt}
\caption{Ablation of SCAN-MCP components. V1 to V3 are leave-one-out variants; V4 is a reference baseline. \#SV and \#TV report confirmed/total reported findings.}
\label{tab:rq1-ablation}
\resizebox{\columnwidth}{!}{%
\begin{tabular}{lccccccc}
\toprule
\textbf{Configuration} & \textbf{Slic.} & \textbf{Filt.} & \textbf{Inter.} & \textbf{\#SV} & \textbf{\#TV} & \textbf{FDR(\%)} & \textbf{Prec.(\%)} \\
\midrule
Full SCAN-MCP & \cmark & \cmark & \cmark & 105/130 & 19/20 & 17.33 & 82.67 \\
\;V1: w/o slicing & \xmark & \cmark & \cmark & 46/215 & 0/0 & 78.60 & 21.40 \\
\;V2: w/o filters & \cmark & \xmark & \cmark & 141/423 & 27/96 & 67.63 & 32.37 \\
\;V3: w/o inter-tool & \cmark & \cmark & \xmark & 105/130 & 0/0 & 19.23 & 80.77 \\
\midrule
\;V4: LLM-only (ref.) & \xmark & \xmark & \xmark & 58/118 & 6/8 & 49.21 & 50.79 \\
\bottomrule
\end{tabular}}
\end{table}

\vspace{2pt}\noindent\textbf{LLM Proposal Filtering.} Table~\ref{tab:rq1-halluc} summarizes proposal rejection and residual error in the available audit logs. The denominators differ by step because each stage produces a different type of intermediate proposal. Tool-handler mapping rejects a small fraction of candidates through local signature or dispatcher checks. Of the 9,871 edge decisions recorded across the 100 sampled servers, 7,785 were rejections (78.87\%). Repeated decisions on the same ordered tool pair were counted separately. PoC synthesis has the largest residual error because exploitability depends on runtime preconditions that may not be available during manual sandbox validation.

\begin{table}[t]
\centering\small
\setlength{\tabcolsep}{4pt}
\caption{LLM proposal filtering and residual error by step.}
\label{tab:rq1-halluc}
\resizebox{\columnwidth}{!}{%
\begin{tabular}{lccc}
\toprule
\textbf{Step} & \textbf{\# Proposed} & \textbf{Rej.(\%)} & \textbf{Resid.(\%)} \\
\midrule
Tool-handler mapping & 668 & ${\approx}$0.4 & 0.0 \\
Edge inference & 9,871 & 78.87 & 1.61 \\
PoC synthesis & 470 & 68.09 & 17.33 \\
\bottomrule
\end{tabular}}
\par\vspace{1pt}
{\footnotesize\raggedright
\emph{Note.} For tool-handler mapping and edge inference, Resid.\ is the complement of component precision (Table~\ref{tab:rq1-stage-metrics}). For edge inference,
$100\% - 98.39\% = 1.61\%$. For PoC synthesis, it is the unconfirmed-over-reported rate under sandbox validation.\par }
\end{table}

\vspace{2pt}\noindent\textbf{Summary.} SCAN-MCP outperforms static and LLM-based baselines on the manually validated 100-server sample. The ablation results show that slicing, source-grounding and pattern filters, and inter-tool reasoning each contribute to precision or coverage, while V3 recovered none of the 19 TVs confirmed by Full on this sample.

\subsection{RQ2: Real-World Vulnerability Reports}
\label{sec:evaluation-c}
\noindent\textbf{Objective.} RQ2 analyzes the full MCP corpus to characterize the distribution and structure of vulnerability reports emitted by SCAN-MCP, with emphasis on reports that require tool-chain reasoning.

\vspace{2pt}\noindent\textbf{Quantitative Overview.}
Across \textit{10,121} MCP servers, SCAN-MCP emits \textit{27,123} reported findings. 
Of these, \textit{8,422} (\textit{31.05}\%) were labeled as tool-chain reports and \textit{18,701} as single-tool reports.
This fraction describes the distribution of SCAN-MCP's reports rather than the prevalence of confirmed vulnerabilities in the ecosystem; we interpret it as an ecosystem-scale risk signal. 

The original TV count deduplicates reports by server, vulnerability category, and ordered tool chain. For single-tool reports, we counted one report per eligible tool-level classification file after removing repeated repository paths. Grouping reports by repository revision, declared final tool, category, and normalized subtype yields 25,511 report groups, including 6,848 tool-chain groups. This metadata grouping measures repetition across reports. Sandbox confirmation establishes the observed effect of a tested path. The number of distinct underlying defects remains unverified.
Table~\ref{tab:rq2-vuln-summary} groups concrete risks such as command execution under Injection and over-broad file access under Broken Access Control.

Of the 20 tool-chain reports on the validation subset, 11 (55\%) came from Python servers and 9 (45\%) came from TypeScript servers. Nineteen of these reports were confirmed. Two-, three-, and four-tool chains accounted for 55\%, 30\%, and 15\% of the validation TV reports, compared with 40.5\%, 33.7\%, and 25.8\% corpus-wide. Among the 20 validation TV reports, 40\% were labeled dataflow-only, 55\% both, and 5\% enablement-only. The corresponding corpus-wide shares were 47.7\%, 44.7\%, and 3.0\%, with 4.6\% undetermined. Servers exposing at least 50 tools contributed 0 of the 20 validation reports and 30.9\% of the corpus tool-chain reports. These coverage gaps limit generalization to other languages and larger servers.

\begin{table}[t]
\centering
\footnotesize
\caption{Distribution of reported findings by category and composition type.}
\label{tab:rq2-vuln-summary}
\begin{tabular*}{\columnwidth}{@{\extracolsep{\fill}}cccc}
\toprule
\textbf{Category} & \textbf{Total} & \textbf{Tool-chain} & \textbf{Single-tool} \\
\midrule
Broken Access Control & 9,111 & 4,178 & 4,933 \\
Injection & 8,669 & 2,285 & 6,384 \\
SSRF & 3,067 & 231 & 2,836 \\
Insecure Design & 3,617 & 1,070 & 2,547 \\
Others & 2,659 & 658 & 2,001 \\
\midrule
\textbf{Total} & \textbf{27,123} & \textbf{8,422} & \textbf{18,701} \\
\bottomrule
\end{tabular*}
\par\smallskip
{\raggedright
\emph{Note.} Reports are counted under the OWASP category recorded in their classification output. \emph{Others} combines the categories that are not listed separately.}\par 
\end{table}

\vspace{2pt}\noindent\textbf{Platform Distribution.} Table~\ref{tab:rq2-platform} shows reported findings by platform. Platform labels are not mutually exclusive because a server may appear in multiple sources. 
MCP.so accounts for the largest report volume. Of the reports associated with the GitHub label, 38.7\% are tool-chain reports.

\begin{table}[t]
\centering
\footnotesize
\caption{Distribution of reported findings by platform.
Platform labels are not mutually exclusive.}
\label{tab:rq2-platform}
\begin{tabular*}{\columnwidth}{@{\extracolsep{\fill}}lccc@{}}
\toprule
\textbf{Platform} & \textbf{\# Repo. paths} & \textbf{\# Findings}
  & \textbf{Tool-chain (\%)} \\
\midrule
MCP.so   & 3,562 & 11,370 & 26.5 \\
Glama    & 2,474 &  7,357 & 31.8 \\
Smithery & 1,997 &  6,040 & 32.0 \\
Pulse    & 2,561 &  7,031 & 32.0 \\
GitHub   & 1,233 &  3,373 & 38.7 \\
\bottomrule
\end{tabular*}
\par\smallskip
{\raggedright
\emph{Note.} Only repository paths with a recorded label
for at least one of the five sources are shown.\par }
\end{table}

\vspace{2pt}\noindent\textbf{Language Distribution.} Table~\ref{tab:rq2-language} shows reported findings by implementation language, with non-exclusive labels for multi-language servers. TypeScript and Python dominate reported-finding volume, while Go had the highest tool-chain fraction (39.3\% of 247 reports from 75 servers). 
The full corpus includes 651 servers labeled JavaScript. The tool-chain fraction in each row is computed over all reports with that language label

\begin{table}[t]
\centering
\footnotesize
\caption{Distribution of reported findings by implementation language. Language labels are not mutually exclusive.}
\label{tab:rq2-language}
\begin{tabular*}{\columnwidth}{@{\extracolsep{\fill}}cccc}
\toprule
\textbf{Language} & \textbf{\# Servers} & \textbf{\# Findings} & \textbf{Tool-chain (\%)} \\
\midrule
TypeScript & 5,771 & 15,364 & 31.6 \\
Python & 4,420 & 11,838 & 33.9 \\
JavaScript & 651 & 2,060 & 28.2 \\
Go & 75 & 247 & 39.3 \\
Others & 131 & 371 & 20.5 \\
\bottomrule
\end{tabular*}
\end{table}

\vspace{2pt}\noindent\textbf{Discussion and Insights.}
Tool-chain reported findings are not isolated anecdotes. Many arise when one tool produces a path, file, state update, or contextual artifact that another tool later treats as trusted input. These findings support the central claim of the paper: hardening individual tools is necessary but insufficient, because unsafe behavior can emerge from the way ordinary tools are composed at run time.

\vspace{2pt}\noindent\textbf{Case Study 1: Arbitrary Command Execution.}
A terminal-style server exposes an \texttt{execute\_command} tool with no input sanitization, allowing an attacker to run arbitrary shell commands. If the server runs with elevated privileges, the attacker can obtain broad host control. We observed this pattern in marketplace-listed servers at analysis time, showing that high-impact single-tool flaws still appear in deployed MCP servers. A sandbox demonstration of this case is available in our artifact~\cite{scanmcp_demo_cmd}.

\vspace{2pt}\noindent\textbf{Case Study 2: Write and Execute Chain.}
In this case, \texttt{execute\_command} blacklists dangerous shell commands and \texttt{write\_file} writes user content to disk. Each tool appears constrained in isolation. By invoking the two tools in sequence, an attacker can write a malicious script through \texttt{write\_file} and execute it through \texttt{execute\_command}, achieving code execution despite the command blacklist. SCAN-MCP identified the chain and confirmed it through manual sandbox validation. A sandbox demonstration of this tool-chain vulnerability is available in our artifact~\cite{scanmcp_demo_write_execute}.

\vspace{2pt}\noindent\textbf{Case Study 3: Browse, Click, and Upload Chain.}
A browser-automation server exposes three relevant tools: \texttt{browser\_navigate}, \texttt{browser\_click}, and \texttt{browser\_file\_upload}. SCAN-MCP identified a candidate chain in which the agent opens an attacker-controlled website with \texttt{browser\_navigate}, uses \texttt{browser\_click} to activate the site's file-selection control, and invokes \texttt{browser\_file\_upload} with a sensitive local path such as \texttt{/etc/passwd}. The first two steps establish the attacker-controlled recipient and file-selection context; upload then supplies the host-local path into that context. In sandbox validation, the selected local file was delivered to the attacker-controlled endpoint, confirming a file-disclosure chain. A sandbox demonstration of this browser-upload chain is available in our artifact~\cite{scanmcp_demo_browser_upload}. We responsibly reported this issue to the vendor of a widely used browser-automation MCP server. The vendor initially considered it outside the project's declared security scope, but about two weeks later patched file-upload path handling. This case illustrates why per-tool review is insufficient: navigation and clicking appear to manipulate only browser state, and upload appears to implement normal file selection; together, they deliver host-local data to an attacker-controlled page.

\subsection{RQ3: Scalability and Cost}
\label{sec:scalability}
\noindent\textbf{Objective.} RQ3 evaluates the scalability of SCAN-MCP for large MCP ecosystems, focusing on runtime, parallel throughput, and LLM cost. We report measured results for Stage~1 tool-handler mapping and Step~III inter-tool reasoning under 10-thread configurations where applicable.

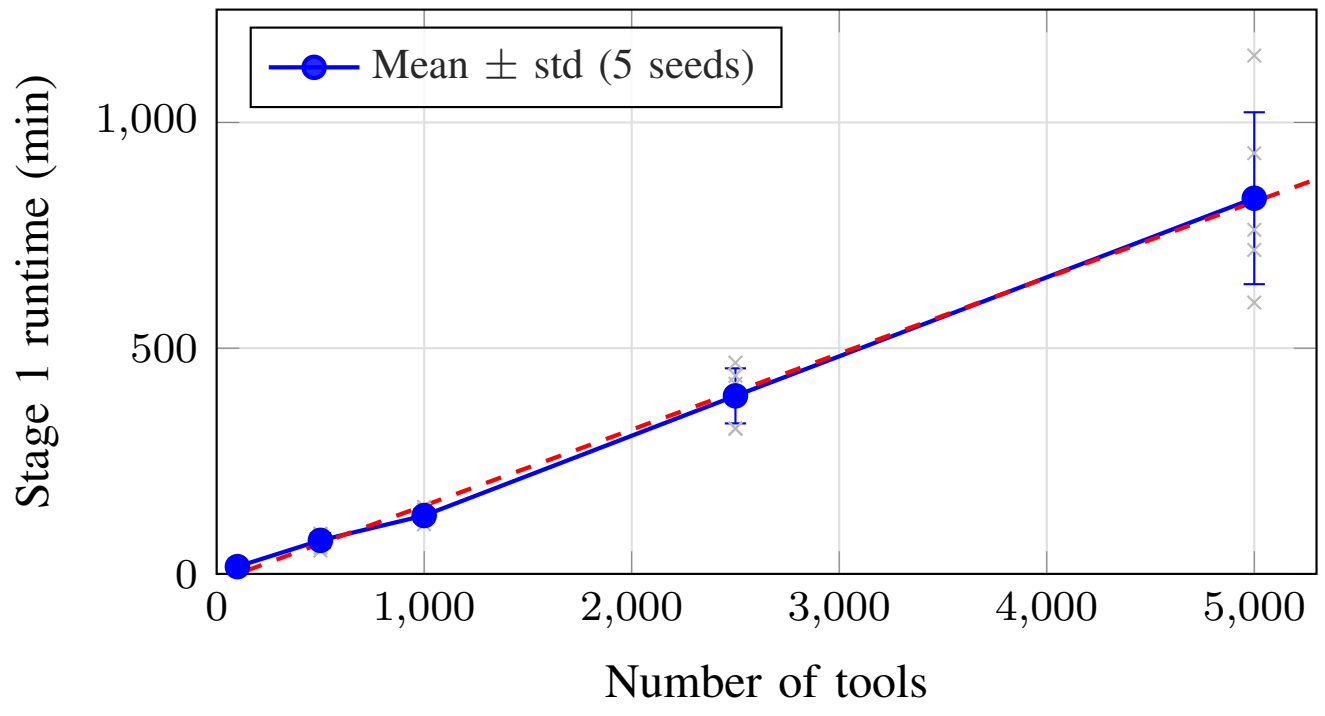
\begin{figure}[t]
\centering
\begin{tikzpicture}
\begin{axis}[
    width=\columnwidth,
    height=0.6\columnwidth,
    xlabel={Number of tools},
    ylabel={Stage~1 runtime (min)},
    xmin=0, xmax=5300,
    ymin=0, ymax=1250,
    grid=major, grid style={gray!25},
    legend pos=north west,
    legend cell align={left},
    legend style={font=\footnotesize, fill opacity=0.85},
    label style={font=\small},
    tick label style={font=\footnotesize},
]
\addplot[only marks, mark=x, mark size=1.8pt, color=gray!55, forget plot]
  coordinates {
    (100,20.5) (100,19.0) (100,19.3) (100,9.6) (100,12.2)
    (500,84.4) (500,58.5) (500,89.5) (500,87.5) (500,51.8)
    (1000,142.7) (1000,121.4) (1000,124.3) (1000,109.9) (1000,148.3)
    (2500,467.9) (2500,421.2) (2500,321.4) (2500,439.0) (2500,322.4)
    (5000,762.2) (5000,601.1) (5000,1148.3) (5000,931.9) (5000,717.7)
  };
\addplot[mark=*, thick, color=blue, error bars/.cd, y dir=both, y explicit]
  coordinates {
    (100,16.1) +- (0,4.4)
    (500,74.3) +- (0,15.9)
    (1000,129.3) +- (0,14.2)
    (2500,394.4) +- (0,61.0)
    (5000,832.2) +- (0,190.4)
  };
\addlegendentry{Mean $\pm$ std (5 seeds)}
\addplot[dashed, thick, color=red, domain=0:5300, forget plot]
  {0.16822*x - 16.89};
\end{axis}
\end{tikzpicture}
\caption{Stage~1 tool-handler mapping runtime under the 10-thread configuration across five tool-count settings, each with five independent seeds. The measured means follow a linear fit of 10.09 seconds per tool with $R^2=0.998$, indicating near-linear scaling up to 5,000 tools.}
\label{fig:scalability}
\end{figure}

\vspace{2pt}\noindent\textbf{Per-Tool Analysis Scalability.}
We measure Stage~1 runtime under a 10-thread configuration at 100, 500, 1,000, 2,500, and 5,000 tools, with five random seeds per setting. Runtime scales nearly linearly: a fit over per-size means gives 10.09 seconds per tool with $R^2=0.998$ (Fig.~\ref{fig:scalability}). This effective wall-clock rate corresponds to about 8,560 tools per day on one 10-thread node, implying an extrapolated Stage~1 runtime of 10.5 days for 90,200 tools and an estimated LLM cost of \$631 at the measured \$0.007 per-tool cost.

\vspace{2pt}\noindent\textbf{Edge and Chain Analysis.}
Step~III evaluates all directed tool pairs in a server, which is $O(n^2)$ in the number of tools $n$. This cost is manageable because most servers expose fewer than 10 tools, giving at most 72 directed pairs per server. On an upper-tail server with $n=22$ tools (462 directed pairs), edge scoring under 10 threads costs 0.44 seconds and $\$4.1\times10^{-4}$ per edge, while candidate-chain scoring costs 1.23 seconds and $\$1.5\times10^{-3}$ per chain. Because Step~III parallelizes over pairs and enumerates only bounded-length simple paths over accepted edges, its overhead remains small relative to single-tool reasoning, which dominates per-tool reasoning cost.

\vspace{2pt}\noindent\textbf{Summary.} SCAN-MCP scales near-linearly with corpus size for Stage~1. The 10-thread scalability experiment supports a measured Stage~1 throughput of approximately 8,560 tools per day, 
while Step~III remains tractable because most MCP servers expose small tool sets and chain enumeration is bounded. The reported tool chains involved two to four tools. 
\section{Discussion}

\vspace{2pt}\noindent\textbf{Defensive Implications.}
SCAN-MCP motivates composition-aware assessment before MCP server deployment. Its intermediate representations, source-grounding requirements, and deterministic gates can inform checks for tool-host capability boundaries, safe parameter propagation, and explicit confirmation before sensitive file, network, or command operations. For each confirmed finding, SCAN-MCP reports the PoC, supporting evidence, and remediation suggestions. Case Study~3 shows why security disclosure boundaries can be unclear for cross-tool findings. The project initially classified our report as out of scope. About two weeks later, an update to the server's file-upload path handling fixed the issue we reported.
Marketplace operators and enterprise security teams could use SCAN-MCP during onboarding and updates, and maintainers could use it in CI. Analysts would prioritize reports and confirm selected effects in isolated test environments. Future work could automate predefined effect checks in resettable environments, with analysts handling deployment-specific setup and ambiguous results.

\vspace{2pt}\noindent\textbf{Limitations.}
First, SCAN-MCP analyzes single-server tool chains and excludes cross-server chains; extending SCAN-MCP to cross-server chains is future work because out-of-band sharing via files, HTTP, prompts, or credentials complicates scalability and reproducibility. Second, full-corpus results count reported findings rather than exhaustively confirmed vulnerabilities; precision is estimated from the manually validated 100-server sample, so full-corpus counts should be read as ecosystem-scale risk signals rather than prevalence estimates. 
Third, source-grounding requirements and pattern filters reduce unsupported LLM output but are not a complete semantic proof, so SCAN-MCP may miss handlers that rely on reflection, runtime code generation, framework-specific dispatch, or external services.
Source-visible wrappers and dispatch branches remain analyzable when the library implementation is unavailable. Within one server, repository-visible reads and writes can support hypotheses about dependencies through a remote database. Precompiled internals, unavailable database-side logic, and runtime-only resource identities limit coverage.
Fourth, sandbox validation may fail when credentials, network dependencies, long-running behavior, or environment-specific preconditions are unavailable.
Global recall requires a reference set of exploitable single-tool and tool-chain cases. Manual confirmation covers emitted reports. Omissions in source/sink labeling or dependency inference can leave cases undetected by both Full and V2. A synthetic benchmark with labeled cases could estimate recall for its predefined cases under controlled conditions.

\vspace{2pt}\noindent\textbf{Threats to Validity.}
\emph{Construct:} Confirmation requires observing the effect that was specified in the report before manual validation. Report-level precision uses all 150 reports from the validation subset as its denominator, including those blocked by unmet runtime prerequisites. The full-corpus figures describe the distribution of emitted reports.
\emph{Internal:} LLMs may hallucinate or omit flows, but SCAN-MCP reduces unsupported output through fixed prompts, temperature~0, source-grounding requirements, deterministic gates, pattern filters, and sandbox confirmation for findings counted as confirmed. 
\emph{External and reliability:} The corpus reflects public repository and marketplace listings from a fixed collection window. Tool-chain validation covers only a subset of corpus languages and server sizes. Hosted-model changes can affect handler mappings, inferred relations, and emitted reports across runs. Reproducibility depends on the recorded models, prompts, and evaluation settings.

\section{Related Work}

\noindent\textbf{LLMs for Security.}
Prior work studies LLM privacy, memorization, training-data extraction, agentic misalignment, and cybersecurity applications~\cite{das2025security,carlini2021extracting,carlini2022quantifying,kassem2025alpaca,lynch2025agentic,xu2024large,jaffal2025large}. These studies focus on model behavior or model-centric attack surfaces. In contrast, we study security failures in the software ecosystem exposed to LLM agents, namely MCP servers and their tool interactions.

\vspace{2pt}\noindent\textbf{Program Analysis for Software Security.}
Static slicing, taint tracking, symbolic execution, and interprocedural frameworks such as IFDS, IDE, and FlowDroid are well established for analyzing source-level dependencies~\cite{reps1995precise,arzt2014flowdroid,thome2015security,jovanovic2010static,baldoni2018survey}. SCAN-MCP uses program-analysis machinery for handler recovery and slicing, but MCP composition introduces runtime edges selected by an LLM or user between independently registered tools. We add inter-tool graph construction, semantic proposals, deterministic gates, and sandbox confirmation.

\noindent\textbf{MCP Security.}
Recent work studies MCP security through taxonomies, benchmarks, ecosystem datasets, server-side detectors, and workflow hardening~\cite{hasan2025model,hou2025model,yang2025mcpsecbench,radosevich2025mcp,lin2025large,wang2025mcpguard,ntousakis2025securing,guo2025systematic}. Closest to our focus, Les Dissonances~\cite{li2025dissonances} studies cross-tool harvesting and polluting in multi-tool LLM agents, showing how tool descriptions and tool interactions can redirect agent workflows or leak information across tools. Mind Your Server~\cite{zhao2025mind} studies parasitic toolchain attacks in the MCP ecosystem, characterizing unintended privacy disclosure patterns that arise when malicious or risky tools are composed with other tools. These works establish that tool composition is a real attack surface. SCAN-MCP is complementary but differs in goal: it targets automated detection and measurement rather than constructing or characterizing one attack family. It also differs in mechanism: SCAN-MCP performs repository-level handler recovery and slicing, constructs inter-tool dependency graphs, and distinguishes reported findings from sandbox-confirmed findings across 10,121 servers and 90,200 tools. Practical scanners such as snyk-agent-scan and mcp-shield target deployed posture, permissions, or policy violations, while mcp-scanner provides a source-level behavioral analyzer that we evaluate directly.
\section{Conclusion}
We presented SCAN-MCP, an ecosystem-scale framework for finding composition-dependent security risks in MCP servers. 
SCAN-MCP recovers handlers and code slices, then uses LLM-assisted proposals and deterministic structural and type checks to generate single-tool and tool-chain reports. Across 10,121 servers with 90,200 discovered tools, SCAN-MCP emitted 27,123 reports, including 8,422 tool-chain reports. On a subset of 100 servers, manual sandbox validation confirmed 124 of 150 reports (82.67\%). Among the 19 confirmed tool-chain findings, sink-only controls reproduced the asserted effect in 0 cases under the tested conditions. These cases illustrate why evaluating MCP tools can require reasoning about their interactions and shared state.

\section*{Acknowledgment}
This work was supported in part by a start-up grant from Nanyang Technological University (Award No.~025559-00001) and by the CyberSG R\&D Programme Office (Award No.~027117-00001, funder reference CRPO-GC6-NTU-001). The UCLA authors are partially supported by Amazon, Cisco, Coefficient Giving, and Keysight.

The authors used OpenAI's ChatGPT/Codex~\cite{visualstudio-code} to assist with drafting and revising text throughout the manuscript. The authors reviewed and edited all AI-assisted text and take full responsibility for the paper's content, claims, citations, and results.

\bibliographystyle{IEEEtran}
\bibliography{reference}

\end{document}